\documentclass[twocolumn,aps]{revtex4}
\usepackage{graphicx}
\usepackage{dcolumn}
\usepackage{amsmath}
\usepackage[latin1]{inputenc}
\begin{document}

\title{Ex-situ control of fine-structure splitting and excitonic binding energies in single InAs/GaAs quantum dots}

\author{R. Seguin}
\author{A. Schliwa}
\author{T. D. Germann}
\author{S. Rodt}
\author{K. P\"otschke}
\author{U. W. Pohl}
\author{D.~Bimberg}
\affiliation{Institut f\"ur Festk\"orperphyik, Technische Universit\"at Berlin, Hardenbergstr. 36, 10623 Berlin}

\begin{abstract}
A systematic study of the impact of annealing on the electronic properties of single InAs/GaAs quantum dots (QDs) is presented. We are able to record single QD cathodoluminescence spectra and trace the evolution of one and the same QD over several steps of annealing. A systematic reduction of the excitonic fine-structure splitting is reported. In addition the binding energies of different excitonic complexes change dramatically. The results are interpreted in terms of a change of electron and hole wavefunction shape and mutual position.
\end{abstract}

\maketitle

Self-assembled quantum dots (QDs) are building blocks for numerous novel devices including single photon emitters and storage devices \cite{benson}. It is of largest importance to tailor their opto-electronical properties for optimal device performance. 
Annealing can considerably alter the electronic structure of QDs \cite{young, langbein, tarta}. Here, the first systematic study of the influence of such an annealing process on the emission characteristics of {\it one and the same} QD for two consecutive steps of annealing is presented. Excitonic binding energies and fine-structure splittings are determined. 

\begin{figure} \label{figure1}
  \includegraphics[width=.99\columnwidth]{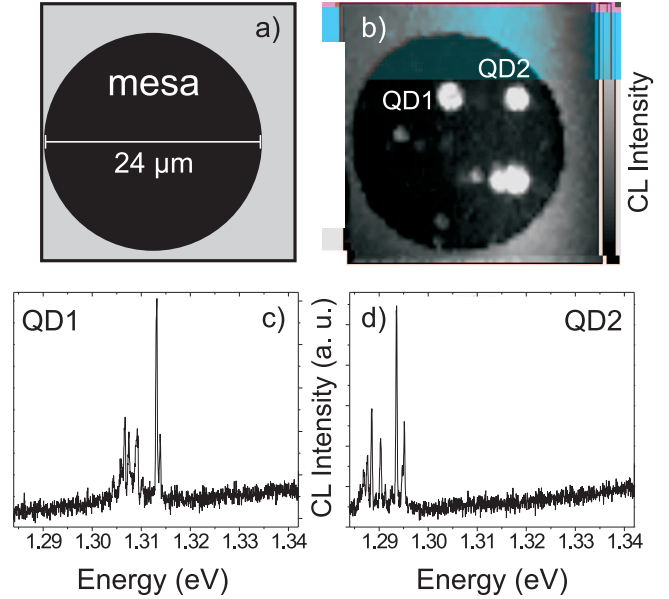}
  \caption{a) Schematic top view of the mesa structure. b) Maxima of monochromatic intensity between 1.284 and 1.342 eV visualize the position of the QDs. Four QDs are located in this particular mesa, corresponding to a QD density of 10$^6$ QDs per cm$^2$ in this spectral range. \mbox{c) / d)} Spectra of QDs 1 and 2.}
\end{figure}

The InAs QDs were grown by MOCVD in GaAs matrix on GaAs(001) substrates. 
For the QDs nominally 1.9 monolayers of InAs were deposited followed by a 540 s growth interruption.  
During the growth interruption the QDs undergo a ripening process \cite{potschke}. Due to its long duration, most QDs gain in size leading to an ensemble peak centered at 1.06 eV. However, some small QDs remain as they represent the material reservoir for the ripening process of the larger QDs. This leads to an ultra-low QD density ($<10^7$ per cm$^2$) in the 1.25-1.35 eV spectral range \mbox{(Fig.\ \ref{figure1})}. 

The sample was examined using a JEOL JSM 840 scanning electron microscope equipped with a cathodoluminescence setup providing temperatures as low as 6 K. The luminescence was dispersed by a 0.3 m monochromator equipped with a 1200 lines/mm grating. The light was detected with a liquid-nitrogen cooled Si charge-coupled-device camera. The minimal linewidth as given by the setup was $\approx$140 $\mu$eV. Using a lineshape analysis, the energetic position of a single lines could be determined withtin an accuracy better than \mbox{20 $\mu$eV}.

In order to relocate the QDs after annealing, circular mesas with 24 $\mu$m in diameter were etched into the sample surface (Fig.\ \ref{figure1}). The consecutive annealing steps lasted five minutes at 710 and 720 $^{\circ}$C respectively, performed under As atmosphere in order to stabilize the sample surface.

\begin{figure} \label{figure2}
  \includegraphics[width=.99\columnwidth]{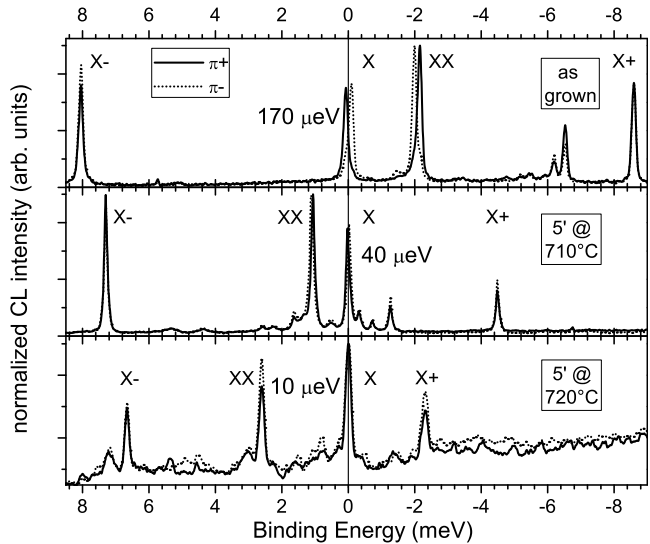}
  \caption{The effect of two annealing steps on the spectrum of a single QD is shown. 0 eV corresponds to the respective exciton recombination energy (1.2738 eV for as grown, 1.3002 for 710 $^{\circ}C$, and 1.3174 eV for 720 $^{\circ}$C). Additionally, the respective excitonic fine structure splitting is shown to decrease from $170 \pm 20$ ${\mu}$eV to $10 \pm 20$ ${\mu}$eV.}
\end{figure}

Fig. \ref{figure2} shows the influence of the annealing steps on a particular QD. Neutral excitons (X), biexcitons (XX) and charged (positively (X+), negatively (X-)) excitons could be identified following Ref. \cite{rodt05}. For easier comparison the energetic position of the X line has been shifted to 0 meV. Fig.\ \ref{figure2} shows that the XX shifts to lower energies with respect to the X line, changing its character from anti-binding (-2.1 meV) to binding (2.6 meV) with a total change in binding energy $\Delta E_{XX}^{bind} = 4.7$ meV. Likewise, the X+ binding energy increases by $\Delta E_{X^{+}}^{bind}=$ 6.3 meV. The X- on the other hand shows the opposite trend becoming less binding with its binding energy decreasing by $\Delta E_{X^{-}}^{bind}=$ -1.3 meV. Additionally the excitonic fine-structure splitting (FSS) was recorded. For this particular dot it decreased from 170 to $\approx$10 $\mu$eV, i.e. a value below our experimental accuracy. The general trend of decreasing FSS and increasing XX binding energy after annealing has also been observed by Young et al. However, they did not record the spectra of identical QDs before and after annealing \cite{young}.

The binding energies of these complexes are a function of the
wavefunction shape and their mutual position affecting the direct Coulomb
energies and the degree of correlation.
The energy contribution due to correlation $E^{corr}$ foremost depends on the number of bound states and the
sublevel spacing. In our case there is a trade off between the decreasing
number of bound states with annealing ($E^{corr}\downarrow$) and
the slightly decreasing sublevel spacing ($E^{corr}\uparrow$). Therefore
we approximate $E^{corr}$ to be constant during annealing and analyze the change of binding energies in terms of the direct Coulomb integrals and their change alone:
\begin{eqnarray}
\Delta E_{X^{+}}^{bind} & = & \Delta J_{eh}+\Delta J_{hh}+(\Delta E_{X^{+}}^{corr}=0)\quad,\label{eq:1}\\
\Delta E_{X^{-}}^{bind} & = & \Delta J_{eh}+\Delta J_{ee}+(\Delta E_{X^{-}}^{corr}=0)\quad,\label{eq:2} \\
\Delta E_{XX}^{bind} & = & 2\Delta J_{eh}+\Delta J_{ee}+J_{hh}+(\Delta E_{XX}^{corr}=0),\label{eq:3} 
\end{eqnarray}

where $J_{ab}$ describes the Coulomb energy between the wavefunctions $\psi_{a}$ and $\psi_{b}$. 

The left hand values are taken from experiment. As a first approximation, the electron wavefunction does not change with annealing due to its small effective mass, leading to the additional assumption $\Delta J_{ee}=0$. Since the right hand side
of the equation system \ref{eq:1}-\ref{eq:3} has rank two only, we can solve eqs.\ \ref{eq:1} and \ref{eq:2} and use eq.\ \ref{eq:3} as a test. Eqs. \ref{eq:1} and \ref{eq:2} yield $\Delta J_{hh}=7.6$ meV and $\Delta J_{eh}=-1.3$ meV respectively. These values are well confirmed by eq.\ \ref{eq:3}.
$J_{hh}$ describes the Coulomb repulsion of the spin-degenerate hole
groundstates and has therefore a negative value. A positive $\Delta J_{hh}$
hence is a sign of an extension of the hole groundstate wavefunction
upon annealing. $J_{eh}$ describes the the Coulomb attraction between
electron and hole groundstate having a positive value. For the $\Delta J_{eh}$
one would expect a value half as large as $-\Delta J_{hh}$ since the hole
wavefunction increases its extent and the electron extent remains virtually
unchanged. But this is only true if electron and hole preserve their
mutual position and their shapes. In contrast, our results can be understood
if we assume that both wavefunctions are originally oriented along orthogonal
directions like $[110]$ and $[1\bar{1}0]$ and loose this misorientation
during annealing. In an elongated QD electron and hole wavefunctions are aligned into the direction of the elongation. Hence the large FSS plus the required misorientation of electron
and hole wavefunction point at an interface-mediated anisotropy resulting
from the lack of inversion symmetry of the underlying zinc-blende
lattice. Annealing destroys the clearly defined interfaces and the confinement anisotropy vanishes. Model calculations show, that piezoelectric fields are insensitive to annealing.

In conclusion, we have recorded emission spectra of single QDs and followed their evolution under an annealing procedure. We have shown, that it is possible to alter the electronic structure of the QDs on the order of meV in a controlled manner. Our results can be understood by a change of electron and hole wavefunction shape and mutual position. We have thus demonstrated a powerful tool to tailor single QDs' electronic properties for their use in potential applications.

This works was supported by the DFG via SfB 296 and the SANDiE Network of Excellence of the European Commision, Contract No. NMP4-CT-2004-500101.

\bibliographystyle{aipproc}

\end{document}